\documentclass[11pt]{article}
\usepackage[english,activeacute]{babel}
\usepackage{epsfig}
\usepackage{amsmath}
\usepackage{epic}
\usepackage{amssymb}
\usepackage{fancybox}
\usepackage{wrapfig}
\usepackage[font=footnotesize,labelsep=period]{caption}
\usepackage{amsthm}
\usepackage{float}
 \usepackage[linktocpage=true,colorlinks]{hyperref}
 \usepackage[capitalise]{cleveref}
\usepackage{xcolor}
\definecolor{dark-red}{rgb}{0.6,0.15,0.15}
\definecolor{dark-blue}{rgb}{0.15,0.15,0.8}
\definecolor{medium-blue}{rgb}{0,0,0.6}
\hypersetup{
    colorlinks, linkcolor={dark-red},
    citecolor={dark-blue}, urlcolor={medium-blue}
}
\allowdisplaybreaks
\usepackage{booktabs}

\usepackage{wrapfig}
\usepackage{cleveref}

\newcommand{\dashedrightarrow}[2][]{\ext@arrow 0359\rightarrowfill@@{#1}{#2}}

\usepackage[all]{xy}
\usepackage{xypic, color, graphics}

\theoremstyle{plain}

\theoremstyle{definition}

 \def\v #1{\vert #1\vert}             %Para denotar elgrado de #1
 
 \def\m #1 #2{(-1)^{{\v #1} {\v #2}}} %Para denotar el signo (-1)^...
 \let \m=\medskip

\def\bea{\begin{eqnarray}}
 \def\eea{\end{eqnarray}}
\newcommand{\beq}{\begin{eqnarray}}
\newcommand{\eeq}{\end{eqnarray}}
\newcommand{\ba}{\begin{array}}
\newcommand{\ea}{\end{array}}
\newcommand{\be}{\begin {equation}}
\newcommand{\ee}{\end{equation}}
\def\picture #1 by #2 (#3){
  \vbox to #2{
    \hrule width #1 height 0pt depth 0pt
    \vfill
    \special{picture #3} % this is the low-level interface
    }
  }
\def\scaledpicture #1 by #2 (#3 scaled #4){{
  \dimen0=#1 \dimen1=#2
  \divide\dimen0 by 1000 \multiply\dimen0 by #4
  \divide\dimen1 by 1000 \multiply\dimen1 by #4
  \picture \dimen0 by \dimen1 (#3 scaled #4)}
  }
  
\usepackage{fancyhdr}
\usepackage[Bjornstrup]{fncychap}
\usepackage{fancyhdr}
\usepackage{appendix}
\usepackage[nottoc,numbib]{tocbibind}
\begin{document}

\centerline{\Large \bf Quasi-algorithmical construction}\vskip 0.25cm
\centerline{\Large \bf of reciprocal transformations}\vskip 0.25cm

\medskip
\medskip

\centerline{C. Sard\'on}
%\medskip
\vskip 0.5cm
\centerline{ICMAT, CSIC, Campus Cantoblanco}
%\medskip
\vskip 0.2cm
\centerline{C/ Nicol\'as Cabrera, 13--15, 28049, Madrid.}
%\medskip

%\medskip

\vskip 1cm

\begin{abstract}
Reciprocal transformations mix the role of the dependent and independent
variables to achieve simpler versions or even linearized versions of nonlinear PDEs.
These transformations help in the identification of a plethora of PDEs available in the Physics and Mathematics
literature. Two different equations, although seemingly unrelated, happen to be equivalent versions of a same equation after a reciprocal transformation.
In this way, the big number of integrable equations could
be greatly diminished by establishing a method to discern which equations are disguised versions of a common underlying problem. 
Then, a question arises: {Is there a way to identify different versions of an underlying common nonlinear problem?} 
Other uselful applications of reciprocal transformations are subsequently discussed and illustrated with examples.\\
%{\bf Please note that if you don't like to submit a paper you only have to submit an Abstract by using the Electronic Submission Form from the Conference web site}.

\end{abstract}

\section{Introduction}

{\it Reciprocal transformations} are a type of transformation
that exchange the role of dependent and independent variables \cite{ClarkFokasAblo,EstPrada2}. 
When the role of dependent and independent variables is switched, the final space of independent variables is called a {\it reciprocal space}. 
As a physical interpretation, whereas the independent variables play the role of positions in the reciprocal space, this number is increased by turning
certain dependent variables into independent variables and viceversa \cite{ConteMusette}.
In particular, in the case of reciprocal transformations, the existence of conserved quantities for their quasi-algorithmical proposal is fundamental \cite{estevez09,estevez51,EstSar,EstSar2,rogers4,rogers5,RogersCarillo}.
These conservation laws are present in models of real physical processes, where reciprocal transformations are applicable.

During the past decades, more attention has been laid upon reciprocal transformations, due to their manageability, quasi-algorithmical way of approach \cite{estevez09,EstSar,EstSar2}.
Indeed, the interest in the topic is reflected in a growing number of works \cite{Abenda,AbendaGrava2,estevez09,EstSar,EstSar2,Ferapontov1,FerapontovPavlov,FerapontovRogersSchief,h00}.
For example, reciprocal transformations were proven to be a useful instrument to transform equations
with peakon solutions into equations that are Painlev\'e integrable \cite{DDH,h00}.
In 1928, the invariance of nonlinear gas dynamics, magnetogas dynamics and general hydrodynamic systems under reciprocal transformations
was extensively studied \cite{Ferapontov1,RogersKingstonShadwick}. Stationary and moving boundary problems in soil mechanics and nonlinear heat conduction
have likewise been subjects of research \cite{FerapontovRogersSchief,Rogers11}.
  Applications of the reciprocal transformation in continuum mechanics are to be found in the monographs by Rogers and Shadwick \cite{rogers1}.
These transformations have also played an important role in the soliton theory and between hierarchies of PDEs \cite{DDH,h00}. Furthermore, the invariance of certain integrable
hierarchies under reciprocal transformations induces auto-B\"acklund transformations \cite{EstSar,EstSar2,rogers6,RogersCarillo,rogers3}.
Some of the most representative properties of reciprocal transformations are they
map conservation laws to conservation laws and diagonalizable systems to diagolizable systems, but act nontrivially on metrics and
on Hamiltonian structures. For instance, the flatness property or the locality of the Hamiltonian structure are not preserved, in general \cite{AbendaGrava1,AbendaGrava2}.

Finding a proper reciprocal transformation is usually a very complicated task. Notwithstanding, in the cases that concern us: as it can be
the case of equations in fluid mechanics,
a change of this type is usually reliable.
For systems of hydrodynamic type with time evolution
\begin{equation}
(u_{j})_t=\sum_{l=1}^k v^{j}_{l}(u)(u_l)_{x_i}, \quad \forall 1\leq i\leq n,\quad j=1,\dots,k
\end{equation}
and $v^{j}_{l}(u)$ are infinitely differentiable functions. These evolution equations appear in gas dynamics, hydrodynamics, chemical kinetics, differential geometry and topological
field theory \cite{DubroNovi,Tsarev1}.

% $C^{\infty}({\rm J}^pN_{\mathbb{R}^n})$,
%For example, in the case of evolution equations in Fluid Dynamics, usually fields that represent the height of the wave or its velocity, are turned into a new set of independent variables.

%{\it Reciprocal transformations} share this definition but require the employment of conservative forms and their properties, as we shall see in forthcoming paragraphs .

Another advantage of dealing with reciprocal transformations is that many of the differential equations reported integrable in the bibliography of differential equations, which are considered seemingly different from one another, happen to be related via reciprocal transformations.
If this were the case, two apparently unrelated equations, even complete hierarchies of PDEs, can be tantamount versions of a unique
problem. In this way, reciprocal transformations give rise to a procedure of relating allegedly new equations
to the rest of their equivalent integrable sisters.
It is an attempt to diminish the ``zoo-botanical approach" in which a number of integrable, specially nonlinear PDEs have been cultivated
by slight modifications in parameters that still fit the integrability conditions of algorithmic, algebraic or geometric nature.
One suspected way to classify equivalent equations is to
find a canonical form shared by all those equations that are related via a reciprocal transformation. It is intuitively based on the singular manifold method (SMM) \cite{EstClark,EstPradaP}
and some clues have already been depicted in \cite{mitesis}.

A second significant application of reciprocal transformations is their utility in the identification of differential equations
which are not integrable according to algebraic tests (for example, the Painlev\'e test is one of them) \cite{estevez51,EstPrada2}.
Precisely, our motivation for the study of reciprocal transformations rooted in the study of the Camassa-Holm hierarchy in $2+1$ dimensions \cite{EstLejaSar}. This hierarchy, 
has been known to be integrable for some time and has an associated linear problem. 
Nevertheless, in its most commonly expressed form \cite{EstLejaSar} it is not integrable according to the Painlev\'e test, nor the 
SMM is constructive. Our conjecture is that if an equation is integrable,
there must be a transformation that will let us turn the initial equation into a new one in which the Painlev\'e test is successful and the 
SMM could be applied. From here, a Lax pair could be derived, among many other properties \cite{mitesis}.
%Therefore, discarding any pathology, one should be able to write down a transformation that brings the equation into a suitable form, in which the Painlev\'e test is applicable.

\section{General Setting}
Let us consider a general manifold $N_{\mathbb{R}^n}\simeq \mathbb{R}^k\times \mathbb{R}^n$, where the first $k$-tuple
refers to the dependent variables $u=(u_1,\dots,u_k)\in \mathbb{R}^k$ and the $n$-tuple denotes the independent variables $x=(x_1,\dots,x_n)\in \mathbb{R}^n$.
We denote by ${\rm J}^p(\mathbb{R}^n,\mathbb{R}^k)$ the space of jets of order $p$ on $N_{\mathbb{R}^n}$. 

The space of $p$-jets will be locally coordinatized by
 $$x_{i},u_j,(u_j)_{x_{i_1}},(u_j)_{x_{i_1}^{j_1},x_{i_2}^{j_2}},\dots,(u_j)_{x_{i_1}^{j_1},x_{i_2}^{j_2},x_{i_3}^{j_3},\dots,x_{i_n}^{j_n}}$$
and such that $i=1,\dots,n$, $j=1,\dots,k,$ $j_1+\dots+j_n\leq p.$
% \begin{definition}
%  By {\it conservation law} we understand an expression on of the form
% \begin{equation}
% \frac{\partial \psi_1}{\partial x_{i_1}}+\frac{\partial \psi_2}{\partial x_{i_2}}=0, 
% \end{equation}
% for certain two values of the indices in between $1 \leq i_1,i_2\leq n$ and two scalar functions $\psi_1,\psi_2 \in C^{\infty}({\rm J}^pN_{\mathbb{R}^n})$. 
% \end{definition}
% \noindent
% One can find a more formal definition of conservation law in \cite{BCDKKSTVV}.

Let us consider a general (possibly nonlinear) system of a number $q$ of PDEs on ${\rm J}^p(\mathbb{R}^n,\mathbb{R}^k)$, 
with higher-order derivative of order $p$,
\begin{equation}\label{genpdert}
\Psi^l=\Psi^l\left(x_{i},u_j,(u_j)_{x_{i_1}},(u_j)_{x_{i_1}^{j_1},x_{i_2}^{j_2}},\dots,(u_j)_{x_{i_1}^{j_1},x_{i_2}^{j_2},x_{i_3}^{j_3},\dots,x_{i_n}^{j_n}}\right),
\end{equation}
for all $l=1,\dots,q$ and $i=1,\dots,n$, $j=1,\dots,k,$ $j_1+\dots+j_n\leq p.$ The notation accords to the usual: $(u_j)_{x_{i_1}}=\partial u_j/\partial x_{i_1}$, and 
this definition is extensible to higher-order derivatives.

Suppose that we know a number $``n"$ of conserved quantities that are expressible in the following form
\begin{equation}
\begin{gathered}\label{conservedq}
A^{(j)}_{x_{i}}\left(x_{i_1},u_j,(u_j)_{x_{i_1}},(u_j)_{x_{i_1}^{j_1},x_{i_2}^{j_2}},\dots\right)=A^{(j')}_{x_{i'}}\left(x_i,u_j,(u_j)_{x_{i_1}^{j_1}},(u_j)_{x_{i_1}^{j_1},x_{i_2}^{j_2}},\dots\right),\\%\nonumber\\
 x_{i}\neq x_{i'},\qquad A^{(j)}\neq A^{(j')},\\
1\leq i\leq n,\qquad j_1+\dots+j_n\leq p, \qquad 1\leq j,j' \leq 2n.
% &A^{(j'')}_{x_{i''}}=A^{(j''')}_{x_{i'''}},\nonumber\\
% &\dots \quad , \quad \dots \quad,\nonumber\\
% &A^{(2n-2 ')}_{x_{i^{2n-2 '}}}=A^{(2n-1 ')}_{x_{i^{2n-1 '}}},
\end{gathered}
\end{equation}
where $A^{(j)}_{x_{i}}=\partial A^{(j)}/\partial x_{i}$, $\smash{A^{(j')}_{x_{i'}}=\partial A^{(j')}/\partial x_{i'}}$ and $A^{(j)},A^{(j')}\in C^{\infty}{\rm J}^p (\mathbb{R}^n, \mathbb{R}^k),$
are different.

If the number of equations in \eqref{conservedq} is equal to the number of indepent variables, we propose a transformation for each  
$\{x_1,\dots,x_n\}$ to a new set $\{z_1,\dots,z_n\}$ as %NC follows
\begin{equation}\label{closed}
dz_{i}=A^{(j)}dx_{i}+A^{(j')}dx_{i'}, \qquad \forall 1\leq i,i'\leq n,\quad \forall 1\leq j,j' \leq 2n.
\end{equation}
such that if the property of closeness is satisfied, %NC edit commas
 $d^2z_{i}=0$ for all $z_{i}$, $i=1,\dots,n$,  we recover the conserved quantities given in \eqref{conservedq}.

\section{Quasi-algorithmical procedure}

We focus on the case in which only {\it one} conserved quantity equation is used. This is due to the number of
physical examples in which a reciprocal transformation based on one single conserved quantity is workable. It implies that only one independent variable is transformed
and it is denoted by $x_{\hat i}$. 

We now search for a function $X(z_1,\dots,z_n)$ such that 
\begin{equation}\label{indintodep}
x_{\hat i}=X(z_{1},\dots,z_{n})
\end{equation}
is turned into a dependent variable.
Simultaneously, we use the conserved quantity to propose the %NC following
 transformation
\begin{equation}
\begin{aligned}\label{rectransf}
&dz_{\hat i}=A^{(j)}dx_{i}+A^{(j')}dx_{i'},\\
&dz_i=dx_i,
\end{aligned}\end{equation} 
$\forall 1\leq i\neq i'\leq n$ and for a fixed value $1\leq \hat{i}\leq n,$
where the independent variables $x_i, \forall i\neq \hat{i}=1,\dots,n$ are untransformed but renamed as $z_i$.

Deriving relation \eqref{indintodep},
\begin{equation}\label{rectransf1}
dx_{\hat i}=\sum_{i=1}^n X_{z_i}dz_i,\quad X_{z_i}=\frac{\partial X}{\partial z_i}
\end{equation}
%for simplicity, we drop the subindex $X_{z_i}$ to $X_{i}$
and by isolating $dz_{\hat i}$ in \eqref{rectransf1}, we have
\begin{equation}\label{rectransf2}
dz_{\hat i}=\frac{dx_{\hat i}}{X_{z_{\hat i}}}-\sum_{i\neq \hat{i}=1}^n \frac{X_{z_{i}}}{X_{z_{\hat i}}}dx_i,
\end{equation}
where we have used that $dz_i=dx_i$ for all $i\neq \hat{i}$ according to \eqref{rectransf}.

Here, by direct comparison of coefficients in \eqref{rectransf} and \eqref{rectransf2}, and if we identify $z_{\hat i}$ with $z_{i_1}$, we have that
\begin{equation}
\begin{aligned}\label{rectransfa}
&A^{(j')}=\frac{1}{X_{z_{\hat i}}},\\
&A^{(j)}=-\frac{X_{z_{i}}}{X_{z_{\hat i}}}.
\end{aligned}\end{equation} 
\noindent
Now we perform the extension of the transformation to higher-order derivatives. In the case of first-order derivatives, it is
\begin{equation}
\begin{aligned}\label{rectransfts}
u_{x_{\hat i}}&=\frac{\partial u}{\partial z_{\hat i}}\frac{\partial z_{\hat i}}{\partial x_{\hat i}}+\sum_{i\neq \hat{i}=1}^n\frac{\partial u}{\partial z_i}\frac{\partial z_i}{\partial x_{\hat i}}=\frac{u_{z_{\hat i}}}{X_{z_{\hat i}}},\\
u_{x_i}&=\frac{\partial u}{\partial z_{\hat i}}\frac{\partial z_{\hat i}}{\partial x_i}+\sum_{i\neq \hat{i}=1}^n\frac{\partial u}{\partial z_i}\frac{\partial z_i}{\partial x_i}=-\frac{X_{z_i}}{X_{z_{\hat i}}}u_{z_{\hat i}}+u_{z_i}, \quad \qquad \forall i\neq \hat{i}.
\end{aligned}
\end{equation} 
This process is recursively applied to achieve higher order derivatives.
In this way, using expressions in \eqref{rectransfa}, \eqref{rectransfts}, etc., we transform a system \eqref{genpdert} with initial variables $\{x_1,\dots,x_n\}$ into 
a new system written in variables $\{z_{1},\dots,z_{n}\}$ and scalar fields $u_j(z_{1},\dots,z_{n}),\, \forall j=1,\dots,k.$ 

From \eqref{rectransfa}, we can extract expressions for
\begin{equation}
z_i,\quad u_j,\quad (u_j)_{z_{i_1}},\quad (u_j)_{z_{i_1}^{j_1},z_{i_2}^{j_2}},\quad \dots,\quad (u_j)_{z_{i_1}^{j_1},z_{i_2}^{j_2},\dots,z_{i_n}^{j_n}}\end{equation} 
for $j_1+\dots+j_n\leq p$, if possible, given the particular form of $A^{(j)}$, $A^{(j')} \in C^{\infty}{\rm J}^p(\mathbb{R}^n,\mathbb{R}^k)$, in each case.

Bearing in mind expression \eqref{indintodep}, the transformation of the initial system \eqref{genpdert} will then read
\begin{equation}
\Psi^l=\Psi^l\left(z_i,X,X_{z_{i}},X_{z_{i_1}^{j_1},z_{i_2}^{j_2}},\dots,X_{z_{i_1}^{j_1},z_{i_2}^{j_2},z_{i_3}^{j_3},\dots,z_{i_n}^{j_n}}\right)
\end{equation}
for all $l=1,\dots,q$ such that $j_1+\dots+j_n\leq p$ and $z_i=z_1,\dots,z_n.$

\section{Example I: application to PDEs}

\subsection{The $n_0$ equation}
The $n_0$ equation \cite{estevez09} defined on ${\rm J}^4(\mathbb{R},\mathbb{R}^3)$ has the form
\begin{equation}
\left(H_{x_1,x_1,x_2}+3H_{x_2}H_{x_1}+n_0\frac{H_{x_1x_2}^2}{H_{x_2}}\right)_{x_1}=H_{x_2x_3}
\end{equation}
and has proven to be integrable in the particular cases in which $n_0=0,-3/4.$ 
For these two cases, a Lax pair formulation was derived in \cite{Abenda,AbendaGrava1}.
It reads
\begin{align}
&\phi_{x_1x_1x_1}-\phi_{x_3}+3H_{x_1}\phi_{x_1}-\frac{k-5}{2}H_{x_1x_1}\phi=0,\nonumber\\
&\phi_{x_1x_2}+H_{x_2}\phi+\frac{k-5}{6}\frac{H_{x_1x_2}}{H_{x_2}}\phi_{x_2}=0.
\end{align}
The compatibility condition of this Lax pair $(\phi_{x_1x_1x_1x_2}=\phi_{x_1x_2x_1x_1})$ retrives the $n_0$ equation
for the integrable values
\begin{align}\label{n0eq}
&H_{x_1x_1x_2}+3H_{x_2}H_{x_1}-\frac{k+1}{4}\frac{H_{x_1x_2}^2}{H_{x_2}}=0,\nonumber \\
&\Omega_{x_1}=H_{x_2x_3}
\end{align}
For this equation, there exists a reciprocal transformation that turns system \eqref{n0eq} into a generalizacion of the Vankhnenko \cite{vakh}
and/or the Degasperi--Procesi equations \cite{DDH} to $2+1$ dimensions.

We construct a reciprocal transformation by a change of the form
\begin{align}
d{x_1}&=\alpha(x,t,T)\left(dx-\beta(x,t,T) dt-\epsilon(x,t,T)dT\right),\nonumber\\
x_2&=t,\quad x_3=T
\end{align}
If the imposition of a closed one-form is accomplished, $d^2x_1=0$, the following equations arise
\begin{equation}\label{transf1}
\alpha_t+(\alpha\beta)_x=0,\quad \alpha_T+(\alpha \epsilon)_x=0,\quad \beta_T-\epsilon_t+\epsilon \beta_x-\epsilon_x\beta=0
\end{equation}
Now we introduce $H_{x_2}=\alpha(x,t,T)^k$ that leads us to an integrability condition for the number $k$. That is, if $k^2=k+2$ is satisfied, then, 
\begin{equation}
H_{x_1}=\frac{1}{3}\left(\frac{\Omega}{\alpha^k}-k\frac{\alpha_{xx}}{\alpha^3}+(2k-1)\left(\frac{\alpha_x}{\alpha^2}\right)^2\right)
\end{equation}
and using \eqref{n0eq}, 
\begin{equation}\label{transf2}
\Omega_x=-k\alpha^{k+1}\epsilon_x,
\end{equation}
So,
\begin{equation}\label{transf3}
\Omega_t=-\beta \Omega_x-k\Omega\beta_x+\alpha^{k-2}\left(-k\beta_{xxx}+(k-2)\beta_{xx}\frac{\alpha_x}{\alpha}+3k\alpha^k\alpha_x\right).
\end{equation}
The reciprocally-transformed set of equations is $\{\eqref{transf1},\eqref{transf2},\eqref{transf3}\}$. Nonetheless, a more convenient form arises
if we introduce the following definitions
\begin{equation}\label{def1}
A_1=\frac{k_1}{3},\quad A_2=\frac{2-k}{3},\quad M=\frac{1}{\alpha^3}
\end{equation}
Then the integrability condition turns into 
$$k^2=k+2 \rightarrow A_1A_2=0, A_1+A_2=1.$$
Using this definitions, the set $\{\eqref{transf1},\eqref{transf2},\eqref{transf3}\}$ can be rewritten as
\begin{multline}\label{final1}
 A_1M\left(\Omega_t+\beta\Omega_x+2\beta_x\Omega+2\beta_{xxx}+2\frac{M_x}{M^2}\right)+\nonumber\\
+A_2\left(\Omega_t+\beta\Omega_x-\Omega\beta_x-M\beta_{xxx}-M_x\beta_{xx}-M_x\right)=0,
\end{multline} 
\begin{equation*}
A_1\left(\Omega_x+2\frac{\epsilon_x}{M}\right)+A_2\left(\Omega_x-\epsilon_x\right)=0,
\end{equation*}
\begin{align}\label{final2}
M_t=3M\beta_x-\beta M_x,\quad M_T=3M\epsilon_x-&\epsilon M_x,\nonumber \\
 &\beta_T-\epsilon_t+\epsilon \beta_x-\epsilon \beta_x=0.
\end{align}
The transformation can also be applied to the spectral problem and after some direct calculations, we have
\begin{equation}
\psi_{xt}=A_1\left(-\beta \psi_{xx}+\left(\beta_{xx}-\frac{1}{M}\right)\psi\right)+A_2\left(-\beta \psi_{xx}-2\beta_x\psi_x-(1+\beta_{xx})\right),\nonumber \\
\end{equation}
\begin{multline}
\psi_T=A_1\left(M\psi_{xxx}+(M\omega-\epsilon)\psi_x\right)+\\
+A_2\left(M\psi_{xxx}+2M_x\psi_{xx}+\left(M_{xx}+\Omega-\epsilon\right)\psi_x\right)
\end{multline}
where we have set 
$$\phi(x_1,x_2,x_3)=\alpha^{\frac{2k-1}{3}}\psi(x,t,T)$$.
\subsubsection{Reduction}
We reduce set of equations in \eqref{final2} by setting $\epsilon=0,\Omega=a_0$. The system reduces to
\begin{align}
&2MA_1\left(\beta_{xx}+a_0\beta-\frac{1}{M}\right)_x-A_2\left(M\beta_{xx}+a_0\beta+M\right)_x=0,\nonumber\\
&M_t=3M\beta_x-\beta M_x.
\end{align}
The reduction of the Lax pair can be obtained by setting $\psi_T=\lambda \psi$.
In this case, the reduced spectral problem is
\begin{multline}
A_1\left(\psi_{xxx}+a_0\psi_x-\frac{\lambda}{M}\psi\right)+\\
+A_2\left(\psi_{xxx}+2\frac{M_x}{M}\psi_{xx}+\frac{1}{M}(M_{xx}+a_0)\psi_x-\frac{\lambda}{M}\psi\right)=0
\end{multline}
and
\begin{multline}
A_1\left(\lambda \psi_t+\psi_{xx}+\lambda \beta \psi_x+(a_0-\lambda\beta_x)\psi\right)+\\
+A_2\left(\lambda \psi_t+M\psi_{xx}+(\lambda \beta+M_x)\psi_x+(a_0+\lambda \beta_x)\psi\right)=0
\end{multline}

\subsubsection*{Degasperis-Procesi equation}
For the case $A_1=0$ and $A_2=1$, we can integrate the reduction as
\begin{align}
&\beta_{xx}+a_0\beta=\frac{1}{M}+q_0,\nonumber \\
&(\beta_{xx}+a_0\beta)_t+\beta \beta_{xxx}+3\beta_x\beta_{xx}+4a_0\beta\beta_x-3q_0\beta_x=0.
\end{align}
For $q_0=0$ and $a_0=-1$, we retrieve the well-known Degasperis-Procesi equation.
The reduced Lax pair is in accordance with
\begin{align}
&\psi_{xxx}-\psi_x-\lambda(\beta_{xx}-\beta)\psi=0,\nonumber\\
&\lambda \psi_t+\psi_{xx}+\lambda \beta \psi_x-(1+\lambda \beta_x)\psi=0
\end{align}
which is equivalent to the Degasperi--Procesi Lax pair \cite{DDH}.

\subsubsection*{Vakhnenko equation}
For the case $A_1=0$, $A_2=1$ and $a_0=0$, we integrate the equations as
\begin{align}
&\beta_{xx}+1=\frac{q_0}{M},\nonumber\\
&\left((\beta_t+\beta\beta_x)_x+3\beta\right)_x=0
\end{align}
which is the derivative of the Vakhnenko equation \cite{vakh}, whose Lax pair is
\begin{align}
&\psi_{xxx}+\frac{M_x}{M}\psi_{xx}+\frac{M_{xx}}{M}\psi_x-\frac{\lambda}{M}\psi=0,\nonumber\\
&\lambda \psi_t+M\psi_{xx}+(\lambda\beta+M_x)\psi_x+\lambda\beta_x\psi=0.
\end{align}
This was the first time that a Lax pair for the Vakhnenko equation had been identified by one of the present authors \cite{estevez09}.

\section{Example II: application to hierarchies of PDEs}
The forthcoming examples will show the application of this procedure to the particular case in which the number of initial independent variables is 3,
with the identification $x_1=x,x_2=y,x_3=t$.
In order to make things clearer, some slight changes
in the notation of the theory or the dimension of $N_{\mathbb{R}^n}$ shall be altered to fit our concrete examples.
\subsection{The Camassa--Holm hierarchy}

The Camassa--Holm hierarchy in $(2+1)$ dimensions (henceforth CHH$(2+1)$) can be written in a compact form as
\begin{equation}
U_T=R^{-n}U_Y, \label{1}
\end{equation}
where $R$ is the recursion operator defined by the composition of two operators $K$ and $J$
\begin{equation}
R=JK^{-1},\quad K=\partial_{XXX}-\partial_X,\quad J=-\frac{1}{2}\left(\partial_XU+U\partial_X\right).
\label{2}
\end{equation}

The $n$ component of this hierarchy can also be rewritten as a set of PDEs by introducing $n$ dependent fields
$\Omega_{[i]}, (i=1,\dots, n)$ in the following way
% \begin{equation}
% \begin{aligned}
% &U_Y=J\Omega^{[1]}\\
% &J\Omega^{[i+1]}=K\Omega^{[i]},\\
%   &U_T=K\Omega^{[n]}, \label{3}
% \end{aligned}
% \qquad i=1,\dots, n-1,\end{equation} 
% NC or 
\begin{equation}
\begin{aligned}
U_Y&=J\Omega_{[1]},\\
J\Omega_{[i+1]}&=K\Omega_{[i]},\\
  U_T&=K\Omega_{[n]}, \label{3}
\end{aligned}
\qquad i=1,\dots, n-1,\end{equation} 
and by introducing two new fields, $P$ and $\Delta$, related with $U$ as
\begin{equation}
U=P^2,\quad\quad P_T=\Delta_X, \label{4}
\end{equation}
we can write the hierarchy in the form of the following set of equations
\begin{equation}
\begin{aligned} &P_Y=-\frac{1}{2}\left(P\Omega_{[1]}\right)_X,\\
&(\Omega_{[i]})_{XXX}-(\Omega_{[i]})_X=-P\left(P\Omega_{[i+1]}\right)_X,  \\
& P_T=\frac{(\Omega_{[n]})_{XXX}-(\Omega_{[n]})_X}{2P}=\Delta_X.\label{5}
\end{aligned}
\qquad i=1,\dots, n-1,\end{equation} 

%It was shown in \cite{EstPrada2} that \eqref{5} can be reduced to the the negative Camassa-Holm hierarchy under the reduction $\frac{\partial}{\partial t}=0$. The positive flow can be obtained under the reduction $\frac{\partial}{\partial x}=\frac{\partial}{\partial y}$.
According to \eqref{conservedq}, the conservative form of the first two equations
\begin{align}
A^{(1)}=P, x_{1}=Y,\quad A^{(2)}=-&\frac{1}{2}(P\Omega_{[1]})_X, x_{2}=X\qquad \text{or} \nonumber\\
 &A^{(1)}=P,x_{1}=T,\quad A^{(2)}=\Delta, x_{2}=X.
\end{align}
 allows us to define the exact derivative
\begin{equation}
dz_0= P\,dX-\frac{1}{2}P\Omega_{[1]}\,dY+\Delta\,dT. \label{6}
\end{equation}
A reciprocal transformation can be introduced by considering the former independent variable $X$ as a field
depending on  $z_0$, $z_1=Y$ and $z_{n+1}=T$, such that $d^2X=0$. From (\ref{6}) we have
\begin{equation}
\begin{aligned}
dX&= \frac{1}{P}\,dz_0+\frac{\Omega_{[1]}}{2}\,dz_1-\frac{\Delta}{P}\,dz_{n+1},\\
Y&=z_1,\\ 
T&=z_{n+1},\label{7}
\end{aligned}\end{equation} 
If we consider the new field $X=X(z_0,z_1,\dots,z_{n+1})$, by direct comparison
\begin{equation}
\begin{aligned}
&X_{z_0}=\frac{1}{P},\\
&X_{z_1}=\frac{\Omega_{[1]}}{2},\\ \label{8}
&X_{z_{n+1}}=-\frac{\Delta}{P},
\end{aligned}\end{equation} 
where $X_{z_i}=\frac{\partial X}{\partial z_i}$. We can now extend the transformation by introducing a new
independent variable $z_i$ for each field $\Omega_{[i]}$ by generalizing (\ref{8}) as
\begin{equation} X_{z_i}=\frac{\Omega_{[i]}}{2},\qquad i=1,\dots, n.\label{9}\end{equation}
Each  of the former dependent fields $\Omega_{[i]},\,(i=1,\dots, n)$ allows us to define a new dependent variable $z_i$ through  definition (\ref{9}).
It requires some calculation (see \cite{EstPrada2} for  details) but it can be proven that the reciprocal transformation (\ref{7})-(\ref{9}) transforms (\ref{5}) to the following set of $n$ PDEs 
on ${\rm J}^4(\mathbb{R},\mathbb{R}^{n+2})$
\begin{equation}-\left(\frac{X_{z_{i+1}}}{X_{z_0}}\right)_{z_0}=\left[\left(\frac{X_{z_0,z_0}}{X_{z_0}}+X_{z_0}\right)_{z_0}-\frac{1}{2}\left(\frac{X_{z_0,z_0}}{X_{z_0}}+X_{z_0}\right)^2\right]_{z_i}, \label{10}\end{equation}
with $i=1,\dots, n.$ Note that each equation depends on only three  variables $z_0, z_i, z_{i+1}$. This result generalizes the one found in \cite{h00} for the first component of the hierarchy.
The conservative form of (\ref{10}) allows us to define a field $M(z_0,z_1,\dots, z_{n+1})$ such that
\begin{equation}
\begin{aligned}M_{z_i}&=-\frac{1}{4}\left(\frac{X_{z_{i+1}}}{X_{z_0}}\right),\quad\quad i=1,\dots, n,\\
 M_{z_0}&=\frac{1}{4}\left[\left(\frac{X_{z_0,z_0}}{X_{z_0}}+X_{z_0}\right)_{z_0}-\frac{1}{2}\left(\frac{X_{z_0,z_0}}{X_{z_0}}+X_{z_0}\right)^2\right].\label{11}\end{aligned}\end{equation} 
It is easy to prove that each $M_i$ should satisfy the following Calogero-Bogoyanlevskii-Schiff equation (the CBS equation) \cite{cbs,calogero} on ${\rm J}^4(\mathbb{R},\mathbb{R}^{n+2})$
\begin{equation}M_{z_0,z_{i+1}}+M_{z_0,z_0,z_0,z_i}+4M_{z_i}M_{z_0,z_0}+8M_{z_0}M_{z_0,z_i}=0, \label{12}\end{equation}
with $ i=1,\dots, n.$
%  We call a $p$-order {\it quasilinear PDE} to a
% a submanifold of ${\rm J}^p(\mathbb{R}^n,\mathbb{R}^k)$ as
%  \begin{equation}
% (u_j)_{x_i}=g(u)(u_j)_{x_1^{j_1},x_2^{j_2},x_3^{j_3},\dots, x_n^{j_n}}+f\left(u_j,(u_j)_{x_{i}},\dots,(u_j)_{x_{i_1}^{j_1},x_{i_2}^{j_2},x_{i_3}^{j_3},\dots, x_{i_n}^{j_n}}\right), 
% \end{equation}
% with $\frac{\partial g}{\partial u}\neq 0$, the subindices $i_1,\dots,i_n$ can take any value between $1,\dots,n$, the sum  $j_1+j_2+j_3+\dots+j_n\leq p$ and  $i=1,\dots,n$, $j=1,\dots,k$, where $u$ denotes 
% % In \cite{ClarkFokasAblo}, the most general quasilinear equation above is found to be mapped via hodograph transformation into linearizable equations, that is, equations
% solvable in terms of either a linear PDE or a linear integral equation.
% Indeed, Gardner associated the solution of the KdV equation with the time-independent Schr\"odinger equation and showed, using ideas of the direct
% and inverse scattering, that the Cauchy problem for the KdV equation could be solved in terms of a linear equation \cite{Gard1,GGKM2}.
% This novelty is today the well-known IST, leading to numerous solutions in branches of water waves, stratified fluids, Plasma Physics, statistical Mechanics
% and Quantum Field theory \cite{AbloClark,AbloKruskalSegur,ARS2,AS}.
% In this way, hodograph transformations facilitate the solvability of certain PDEs that can either be treated by the IST or by
% transformation to a linear PDE, (those said to be linearizable). 

{The CBS equation has the Painlev\'e property} and the SMM can be successfully used to derive its Lax pair \cite{EstPrada1}. In \cite{EstPrada2} it was proven that the Lax pair of CBS yields the following spectral problem for the CHH(2+1) hierarchy (\ref{3})
 \begin{eqnarray}\label{13}&&\Phi_{XX}+\frac{1}{4}\left(\lambda U-1\right)\Phi=0,\nonumber\\&&\Phi
 _T-\lambda^n\Phi_Y-\frac{\lambda}{2}C\Phi_X+\frac{\lambda}{4}C_X\Phi=0\end{eqnarray}
 where $$C=\sum_{i=1}^n\lambda^{n-i}\Omega_{[i]}$$ and $\lambda(Y,T)$ is a non-isospectral parameter that satisfies
 \begin{equation}\lambda_X=0,\quad \lambda_T-\lambda^n\lambda_Y=0.\label{14}\end{equation}
 Consequently the problems that we meet when we try to apply the Painlev\'e test to CHH(2+1) \cite{gp95} can be solved owing to  the existence of a reciprocal transformation that transforms the CHH(2+1) hierarchy to $n$ copies of the CBS equation, for which the Painlev\'e methods are applicable.

\subsection{The modified Camassa--Holm hierarchy}
In \cite{estevez51}, the modified Camassa--Holm hierarchy in $(2+1)$ dimensions (henceforth mCHH$(2+1)$), was introduced 
\begin{equation}
u_t=r^{-n}u_y, \label{15}
\end{equation}
where $r$ is the recursion operator, defined by two operators $j$ and $k$ and generalizes the hierarchy introduced by Qiao \cite{Qiao1}
\begin{equation}
r=jk^{-1},\quad k=\partial_{xxx}-\partial_x,\quad j=-\partial_x\,u\,(\partial_x)^{-1}\,u\,\partial_x.%\quad \textrm{where}\quad \partial_x=\frac{\partial}{\partial x}.
\label{16}
\end{equation}
% [NC really the first instance of $\partial_x$?]

%We shall  briefly summarize the results of \cite{estevez51} when a  procedure similar to that described for CHH(2+1) is applied to mCHH(2+1).

If we introduce $2n$ auxiliary fields $v_{[i]}$, $\omega_{[i]}$ defined through
% \begin{equation}
% \begin{aligned}
% & u_y=jv^{[1]},\\
% &jv^{[i+1]}=kv^{[i]},\\
% %\quad 
% &\omega_x^{[i]}=uv_x^{[i]},\\
%  & u_t=kv^{[n]}, \label{17}
% \end{aligned}
% \qquad i=1,\dots, n-1,\end{equation} 
% NC or
\begin{equation}
\begin{aligned}
 u_y&=jv_{[1]},\\
jv_{[i+1]}&=kv_{[i]},\\
%\quad 
(\omega_{[i]})_x&=u(v_{[i]})_x,\\
  u_t&=kv_{[n]}, \label{17}
\end{aligned}
\qquad i=1,\dots, n-1,\end{equation} 
the hierarchy can be written as the system
\begin{equation}
\begin{aligned}
u_y&=-\left(u\omega_{[1]}\right)_x,\\
(v_{[i]})_{xxx}-(v_{[i]})_x&=-\left(u\omega_{[i+1]}\right)_x, \label{18}\\
u_t&=(v_{[n]})_{xxx}-(v_{[n]})_x=\delta_x,
\end{aligned}
\qquad i=1,\dots, n-1,\end{equation} 
According to the theory in \eqref{conservedq}, 
\begin{align}
A^{(1)}=u, x_{1}=y,\quad A^{(2)}=-&u\omega_{[1]}, x_{2}=x\qquad \text{or} \nonumber\\
 &A^{(1)}=u,x_{1}=t,\quad A^{(2)}=\delta, x_{2}=x,
\end{align}
which allows to define the exact derivative
\begin{equation}
dz_0= u\,dx-u\omega_{[1]}\,dy+\delta\,dt \label{19}
\end{equation}
and $z_1=y, z_{n+1}=t$.
We can define a reciprocal transformation such that  the former independent variable $x$ is a new field $x=x(z_0,z_1,\dots , z_{n+1})$ depending on $n+2$ variables in the form
\begin{equation}
\begin{aligned}
x_{z_0}&=\frac{1}{u},\\
x_{z_1}&=\omega_{[1]},\label{20}\\
x_{z_{n+1}}&=-\frac{\delta}{u}.
\end{aligned}\end{equation} 
If we introduce the auxiliary variables for the auxiliary fields, $x_{z_i}=\omega_{[i]}$ for $i=2,\dots,n$,
the transformation of the equations in (\ref{18}) yields a system of equations on ${\rm J}^4(\mathbb{R}^{n+2},\mathbb{R}).$ Note that each equation depends on three variables: $z_0, z_i, z_{i+1}$.
\begin{equation}\left(\frac{x_{z_{i+1}}}{x_{z_0}}+\frac{x_{z_i,z_0,z_0}}{x_{z_0}}\right)_{z_0}=\left(\frac{x_{z_0}^2}{2}\right)_{z_i},\qquad i=1,\dots, n. \label{21}\end{equation}

The conservative form of (\ref{21}) allows us to define a field $m=m(z_0,z_1,\dots, z_{n+1})$ such that on ${\rm J}^3(\mathbb{R}^2,\mathbb{R}^{n+2})$ we have
\begin{equation} m_{z_0}=\frac{x_{z_0}^2}{2},\quad m_{z_i}=\frac{x_{z_{i+1}}}{x_{z_0}}+\frac{x_{z_i,z_0,z_0}}{x_{z_0}},\qquad i=1,\dots, n.\label{22}\end{equation}
Equation (\ref{21}) has been extensively studied from the point of view of  Painlev\'e analysis \cite{EstPrada1} and it can be considered as the modified version of the CBS equation (mCBS) (\ref{12}). Actually, in \cite{EstPrada1} it was proven that the {\it Miura transformation} that relates (\ref{12}) and (\ref{22}) is
\begin{equation}
4M=x_{z_0}-m. \label{23}
\end{equation}
A non-isospectral Lax pair  was  obtained for (\ref{22}) in \cite{EstPrada1}. By inverting this Lax pair through the reciprocal transformation (\ref{20}) the following spectral problem was obtained for mCHH(2+1). This Lax pair \cite{EstPrada1} reads 

\begin{align} 
\left(\begin{array}{c} \phi \\ \hat\phi
\end{array}\right)
_x &=\frac{1}{2} \left[\begin{array}{cc} -1& I\sqrt{\lambda}u \\ I\sqrt{\lambda}u
& 1
\end{array}\right]
 \left(\begin{array}{c} \phi \\ \hat\phi
\end{array}\right)
,\label{24}
\\
% \begin{eqnarray} 
\left(\begin{array}{c} \phi \\ \hat\phi
\end{array}\right)
_t &=\lambda^n 
\left(\begin{array}{c} \phi \\ \hat\phi
\end{array}\right)
_y+\lambda a 
\left(\begin{array}{c} \phi \\ \hat\phi
\end{array}\right)
_x+
% \\
I
\frac{\sqrt{\lambda}}{2} \left[\begin{array}{cc} 0& b _{xx}-b_x  \\
b_{xx}+b_x &0
\end{array}\right]
\left(\begin{array}{c} \phi \\ \hat\phi
\end{array}\right)
,
\end{align}
where
 \begin{equation}
a=\sum_{i=1}^n\lambda^{n-i}\omega_{[i]}, \quad b=\sum_{i=1}^n\lambda^{n-i}v_{[i]},\quad I=\sqrt{-1}\quad i=1,\dots,n,
      \end{equation} 
and $\lambda(y,t)$ is a non-isospectral parameter that satisfies
\begin{equation}\lambda_x=0,\quad \lambda_t-\lambda^n\lambda_y=0.\label{25}\end{equation}
Although the Painlev\'e test cannot be applied to mCHH(2+1), reciprocal transformations are a tool that can be used to write the hierarchy as a set of mCBS equations to which  the Painlev\'e analysis (the SMM in particular) can be successfully applied.

\section{Conclusions}

We have first proposed reciprocal transformations for a single PDE. We have found two interesting reductions for the transformed equation of our proposed model.
One corresponds to the Vakhnenko equation, the other is a Degasperis--Procesi type equation. In this way, we can say that our proposed
equation is a generalization to a higher dimension of the mentioned equations.

Secondly, we have proposed reciprocal transformations for complete hierarchies of PDEs.
In particular, we have contemplated the CHH$(2+1)$ and the mCHH$(2+1)$.
We have presented both hierarchies and have discussed some general properties for a scalar field $U(X,Y,T)$
and $u(x,y,t)$ respectively. We have constructed reciprocal transformations that connect both hierarchies with the CBS and mCBS equations, respectively.
% One of the advantages of the reciprocal transformation is transforming initial equations which do not pass the Painlev\'e test, into others
% that are integrable in the Painlev\'e sense. As it is the case of the Camassa--Holm and Qiao hierarchies which, in their initial format, do not
% possess the Painlev\'e property. Notwithstading, the transformed CBS and mCBS equations do possess the Painlev\'e property.
The Lax pair for both of the hierarchies can be retrieved through those of the CBS and mCBS, correspondingly. If we consider
the Lax pair of the CBS and mCBS and undo the reciprocal transformation, we achieve the Lax pair of the hierarchies.
% In this way, we have achieved the Lax pair for the former hierarchies.

As illustrated through the examples, we can say that reciprocal transformations help us reduce the number of available nonlinear equations in the literature, as two
seemingly different equations can be turned from one into another by reciprocal transformation, which is the case of the CHH$(2+1)$, mCHH$(2+1)$ 
with the CBS and mCBS, correspondingly.
A question for future work is whether there exists a canonical description for differential equations. Intuitively, we expect that if two equations are essentially the
same, although apparently different in disguised versions, they must share the same singular manifold equations. But this is just an initial guess worth of further
research in the future. 

Also, reciprocal transformations have proven their utility in the derivation of Lax pairs.
As we know, obtaining Lax pairs is a nontrivial subject. The common way is to impose ad hoc forms
for such linear problems and make them fit according to the compatibility condition. Accounting for reciprocal transformations, we do not
face the problem of imposing ad hoc Ans\"atze.
One initial equation whose Lax pair is unknown, can be interpreted as the reciprocally-transformed equation whose Lax pair is acknowledged. In this way, undoing the transformation on the latter Lax pair,
we achieve the Lax pair of the former. This is precisely the procedure followed along our examples. 

As a last property, to mention that reciprocal transformations permit us to (sometimes) obtain equations integrable in the Painlev\'e sense if they did
not have this property before the change. This is due to the noninvariability of the Painlev\'e test under changes of variables.

Some possible future research on this topic would consist of understanding whether the singular manifold equations can constitute a canonical representation
of a partial differential equation and designing techniques to derive Lax pairs in a more unified way.
Also, the trial of composition of reciprocal transformations with transformations of other nature,
can lead to more unexpected but desirable results.


\begin{thebibliography}{}
\bibitem{Abenda}
S. Abenda,
{Reciprocal transformations and local hamiltonian structures of hydrodynamic type systems,}
{\em J. Phys. A} {\bf 42}, 095298, 2009.


\bibitem{AbendaGrava1}
S. Abenda, T. Grava,
{Modulation of Camassa--Holm equation and reciprocal transformations,}
{\em Ann. Inst. Fourier} {\bf 55}, 1803--1834, 2005.

\bibitem{AbendaGrava2}
S. Abenda, T. Grava,
{Reciprocal transformations and flat metrics on Hurwitz spaces,}
{\em J. Phys. A} {\bf 40}, 10769--10790, 2007.

\bibitem{BCDKKSTVV}
A.V. Bocharov, V.N. Chetverikov {\it et al.},
{Symmetries and conservation laws for differential equations of mathematical physics,}
{\em Translations of mathematical monographs {\bf 182}, American Mathematical Society, Providence, R.I.} 1999.

\bibitem{cbs} 
O.I. Bogoyavlenskii,
 Breaking solitons in $2+1$ dimensional integrable equations,  
{\em Russian Math. Surv.} \textbf{45}, 1--86, 1990.

\bibitem{Calo}
F. Calogero,
A method to generate solvable non-linear evolution equations,
{\em Lett. Nuovo Cimento,} {\bf 14}, 443--448, 1975. 

\bibitem{calogero}
 F. Calogero, 
 Generalized Wronskian relations, one-dimensionall Schr\"odinger equation and non-linear partial differential equations solvable by the inverse-scattering method,
 {\em Nuovo Cimento B}  \textbf{31}, 229--249 , 1976.


\bibitem{ClarkFokasAblo}
P.A. Clarkson, A.S. Fokas, M.J. Ablowitz,
{Hodograph transformations of linearizable partial differential equations,}
{\em SIAM J. of Appl. Math.} {\bf 49}, 1188--1209, 1989.


\bibitem{ConteMusette}
R. Conte, M. Musette,
{The Painlev\'e Handbook,}
{\em Springer \& Canopus Publishing Limited, Bristol}, 2008.

\bibitem{DDH}
A. Degasperis, D.D. Holm, A.N.W. Hone,
{A new integral equation with peakon solutions,}
{\em Theor. Math. Phys.} {\bf 133}, 1463--1474, 2002.

\bibitem{DubroNovi}
B.A. Dubrovin, S.P. Novikov,
{Hydrodynamics of weakly deformed soliton lattices. Differential geometry and Hamiltonian theory,}
{\em Russ. Math. Surv.} {\bf 44}, 35--124, 1989.

\bibitem{estevez09}
 P.G. Est\'evez,
 Reciprocal transformations for a spectral problem in $2+1$ dimensions,  
{\em Theor. and Math. Phys.} \textbf{159}, 763--769, 2009.

\bibitem{estevez51}
 P.G. Est\'evez,  
Generalized Qiao hierarchy in 2+1 dimensions: reciprocal transformations,
spectral problem and non-isospectrality,
 {\em Phys. Lett. A}  {\bf 375}, 537-540, 2011.

\bibitem{EstClark}
P.G. Est\'evez, P. Clarkson,
Discrete equations and the singular manifold method
in {\em Symmetries and Integrability of Difference Equations III},
 Centre de Recherches Mathématiques Proceedings and Lecture Notes Series, {\bf 25},
 American Mathematical Society, Providence, RI, 139-146, 2000. 
\bibitem{EstLeble}
P.G. Est\'evez, S.L. Leble,
A wave equation in $2+1$ dimensions: Painleve analysis and solutions,
{\em Inverse Problems} {\bf 11}, 925--937, 1995.

\bibitem{EstLejaSar}
P.G. Est\'evez, J.D. Lejarreta, C. Sard\'on, 
Non isospectral $1+1$ hierarchies arising from a Camassa--Holm hierarchy in $2+1$ dimensions,
{\em J. Nonlin. Math. Phys.} {\bf 8}, 9--28, 2011.

\bibitem{EstPrada1}
P.G. Est\'evez, J. Prada,
{A generalization of the sine-Gordon equation to $2+1$ dimensions,}
{\em J. Nonlin. Math. Phys.} {\bf 11}, 164--179, 2004.

\bibitem{EstPradaP}
P.G. Est\'evez, J. Prada,
Singular Manifold method for an equation in $2+1$ dimensions,
{\em J. Nonlin. Math. Phys.} {\bf 12}, 266--279, 2005.

\bibitem{EstPrada2}
P.G. Est\'evez, J. Prada,
{Hodograph transformations for a Camassa--Holm hierarchy in $2+1$ dimensions,}
{\em J. Phys. A} {\bf 38}, 1--11, 2005.

\bibitem{EstSar}
P.G. Est\'evez, C. Sard\'on,
Miura reciprocal Transformations for hierarchies in 2+1 dimensions,
{\em J. Nonlin. Math. Phys.} {\bf 20}, 552--564, 2013.

\bibitem{EstSar2}
P.G. Est\'evez, C. Sard\'on,
Miura reciprocal transformations for two integrable hierarchies in $1+1$ dimensions,
{\em Proceedings GADEIS (2012)}, Protaras, Cyprus, 2012.


\bibitem{Ferapontov1}
E.V. Ferapontov,
{Reciprocal transformations and their invariants,}
{\em Differential equations} {\bf 25}, 898--905, 1989.

\bibitem{FerapontovPavlov}
E.V. Ferapontov, M.V. Pavlov,
{Reciprocal transformations of Hamiltonian operators of hydrodynamic type: nonlocal Hamiltonian formalism for nonlinearly degenerate systems,}
{\em J. Math. Phys.} {\bf 44}, 1150--72, 2003.

\bibitem{FerapontovRogersSchief}
E.V. Ferapontov, C. Rogers, W.K. Schief,
{Reciprocal transformations of two component hyperbolic system and their invariants,}
{\em J. Math. Anal. Appl.} {\bf 228}, 365--376, 1998.

\bibitem{gp95} 
C.R. Gilson, A. Pickering,
 Factorization and Painlev\'e analysis of a class of nonlinear third-order partial differential
equations,
{\em J. Phys. A} {\bf 28}, 2871--2878, 1995.

\bibitem{h00}
A.N.W. Hone,
{Reciprocal link for $2+1$ dimensions extensions of shallow water equations,}
{\em Appl. Math. Lett.} {\bf 13}, 37--42, 2000.



\bibitem{rogers6}
W. Oevel, C. Rogers,
{Gauge transformations and reciprocal links in $2+1$ dimensions,}
{\em Rev. Math. Phys.} {\bf 5}, 299--330, 1993.

\bibitem{Qiao1}
Z. Qiao, L. Liu,
A new integrable equation with no smooth solitons,
{\em Chaos Solitons Fract.} {\bf 41}, 587--593, 2009.

\bibitem{Rogers11}
C. Rogers,
{Application of a reciprocal transformation to a two-phase Stefan Problem,}
{\em J. Phys. A} {\bf 18}, L105-L109, 1985.

\bibitem{rogers4}  
C. Rogers, 
Reciprocal transformations in $(2+1)$ dimensions,
 {\em J. Phys. A} \textbf{19}, L491--L496, 1986.

\bibitem{rogers5} 
 C. Rogers, 
The Harry Dym equation in $2+1$ dimensions: A reciprocal link with the Kadomtsev-Petviashvili equation,
 {\em Phys. Lett. A} \textbf{120}, 15--18, 1987.





\bibitem{RogersCarillo}
C. Rogers, S. Carillo,
{On reciprocal properties of the Caudrey, Dodd--Gibbon and Kaup--Kuppersmidt hierarchies,}
{\em Phys. Scripta} {\bf 36}, 865--869, 1987.


\bibitem{RogersKingstonShadwick}
C. Rogers, J.G. Kingston, W.F. Shadwick,
On reciprocal type invariant transformations in magneto-gas dynamics,
{\em J. Math. Phys.} {\bf 21}, 395--397, 1980.

\bibitem{rogers3}
C. Rogers, M.C. Nucci,
{On reciprocal B\"acklund transformations and the Korteweg de Vries hierarchy,}
{\em Phys. Scripta} {\bf 33}, 289--292, 1986.


\bibitem{rogers1}
C. Rogers, W.F. Shadwick,
{ B\"acklund transformations and their applications,}
{\em Academic Press, New York,} 1982.

\bibitem{mitesis}
C. Sard\'on,
{\em Lie systems, Lie symmetries and reciprocal transformations}


\bibitem{Tsarev1}
S.P. Tsarev,
{The geometry of Hamiltonian systems of hydrodynamic type: the generalized hodograph method,}
{\em Math. USSR Izv.} {\bf 37}, 397--419, 1991.

% \bibitem{vino}
% A. Vinogradov,
% {What are symmetries of nonlinear PDEs and what are they themselves?}

\bibitem{vakh}
V.A. Vakhnenko,
Soltons in a nonlinear model medium,
{\em J. Phys. A} {\bf 25}, 4181--4187, 1992.
% % \bibitem{DJM2005}
% % G. DAmico, J. Janssen and R. Manca. Homogeneous semi-Markov reliability models for credit risk management. {\em Decisions in Economics and Finance}, 28, 2, 79--93, 2005.
% % \bibitem{JM2007}
% % J. Janssen and R. Manca. {\em Semi-Markov risk models for finance, insurance and reliability}, Springer, New York, 2007.
% % \bibitem{JS1995}
% % J. Janssen and C.H. Skiadas. Dynamic modelling of Life-Table data, {\em Applied Stochastic Models and Data Analysis}, 11, {\bf 1}, 35--49, 1995.
% % \bibitem{GS1990}
% % G. Saporta. {\em Probabilit\'es, Analyse des Donn\'ees et Statistique}, Editions Technip, Paris, 1990.
% % \bibitem{CHS2009}
% % C. H. Skiadas and C. Skiadas. {\em Chaotic Modeling and Simulation: Analysis of Chaotic Models, Attractors and Forms}, Taylor and Francis/CRC, London, 2009.
\end{thebibliography}
\end{document}